\documentclass[prd, onecolumn, floatfix, letterpaper, nofootinbib, amsmath, amssymb]{revtex4}

\usepackage{graphicx}
\usepackage{color}
\usepackage{amsmath}
\usepackage{amssymb}
\usepackage{enumerate}
\usepackage{graphics,epsfig,subfigure}
\usepackage{epstopdf}
\usepackage{color}

\newcommand\be{\begin{equation}}
\newcommand\ee{\end{equation}}
\newcommand\ba{\begin{eqnarray}}
\newcommand\ea{\end{eqnarray}}

\begin{document}

\title{Reheating and Entropy Perturbations in Fibre Inflation}

\author{Bao-Min Gu$^{1,2}$\footnote{gubm15@lzu.edu.cn} and Robert Brandenberger$^{2}$\footnote{rhb@physics.mcgill.ca}}

\affiliation{$^1$Institute of Theoretical Physics \& Research Center of Gravitation, Lanzhou University, Lanzhou 730000, China}
\affiliation{$^2$ Department of Physics, McGill University, Montr\'{e}al, QC, H3A 2T8, Canada}

\begin{abstract}

We study reheating in some one and two field realizations of {\it Fibre Inflation}. We find
that reheating begins with a phase of preheating in which long wavelength fluctuation
modes are excited. In two field models there is a danger that the parametric amplification
of infrared fluctuations in the second scalar field - associated with an entropy mode -
might induce an instability of the
curvature fluctuations. We show that, at least in the models we consider, the  entropy mode
has a sufficiently large mass to prevent this instability. Hence, from the point of view of
reheating the models we consider are well-behaved.

\end{abstract}

\pacs{04.50.Kd, 98.80.-k}

\maketitle

\section{Introduction}

The reheating phase (see e.g. \cite{ABCM, Karouby} for recent reviews) is an integral
part of any successfull inflationary model.
The inflationary phase leaves behind a state with an exponentially
suppressed density of matter particles \footnote{Warm inflation \cite{Berera} is an
exception as in this scenario matter particles are produced continuously and
efficiently throughout the inflationary phase, and there is no need for a separate
period of reheating.}, and a mechanism is required which transforms the
energy trapped in the homogeneous inflaton \footnote{The inflaton is the scalar
field whose potential energy drives the inflationary expansion.} condensate to quanta of the regular
matter fields.

Initially, reheating was studied perturbatively \cite{AFW, DL, Stein}, but this neglects
the coherent nature of the inflaton condensate. It was then realized \cite{TB, DK}
that the reheating phase may begin with a period of parametric resonance instability called
{\it preheating} \cite{KLS1, STB, KLS2} during which the oscillations of the inflaton
condensate lead to exponential increase in the number density of long wavelength
fluctuations of either the inflaton field itself (``self-resonance'') or of fields which are
coupled to the inflaton. The resulting state of matter after preheating has a non-thermal
distribution, and hence a second stage of reheating is required in order to obtain
full kinetic and chemical equilibrium.

The equation of state of matter becomes radiation-dominated even after an efficient
preheating period. Since the detailed predictions of any inflationary model for
observations depends on the length of the period after inflation before the
onset of radiation-domination \footnote{The number of e-foldings before the end
of inflation when fluctuations on a given physical scale today exit the Hubble
radius and hence the inflationary slow-roll parameters evaluated at that scale
depend on the length of the reheating phase before the onset of radiation-domination.}
it is important in any inflationary model to study the possible presence of
parametric resonance.

One might fear that the exponential increase in infrared matter fluctuations
during preheating might lead to an exponential increase of the cosmological
fluctuations \cite{BKM} \footnote{Note that such a process would not violate
causality \cite{FB1} since we are talking about modes which, although must
larger than the Hubble radius $H^{-1}$ at the time of reheating, are smaller
than the horizon.}. In single matter field models, however, this does not
happen since the curvature fluctuations are conserved on super-Hubble
scales, as can be shown even beyond perturbation theory (see e.g.
\cite{LV, AB}). However, in models in which there are extra light fields in addition to the inflaton, there
is the danger than fluctuations in the extra fields will also experience parametric
resonance of infrared modes \cite{FB2, BV}. If the fluctuations of these entropy
modes are coupled to the curvature fluctuations, then the exponentially amplified
entropy fluctuations could induce exponentially enhanced curvature fluctuations -
on scales which are measured today, hence destroying the agreement between
theory and observations. Toy models in which such a dangerous amplification of
entropy modes occurs were studied in \cite{us}.  On the other hand, it was shown
that a number of string-motivated inflationary models such as D3-D7 brane
inflation \cite{D3D7} and axion monodromy inflation \cite{axion} are safe
from this potential problem.

Fibre inflation \cite{Cicoli} is a popular model of inflation motivated
by ideas from string theory \footnote{We are not addressing here the possibility
that this model lies in the {\it swampland} \cite{swamp} and it not consistent
with principles of superstring theory \cite{Vafa}.}. In this paper we show that
in the versions of the scenario which we consider here, reheating begins with
a period of preheating \footnote{The possibility of preheating in fibre inflation was also considered in \cite{Antusch}, and the macro reheating properties in \cite{Cabella:2017zsa}.}. On the other hand, we show that the entropy modes
are sufficiently heavy such that no efficient resonance of these modes occurs
during the phase of preheating. Hence, it appears that from the point of view
of reheating constraints, the fibre inflation model is safe.

A word on our notation: we use natural units in which the speed of light
and Planck's constant are set to $1$. Unless otherwise indicated we work
in units in which the Planck mass is also set to $1$. We work in the context of a spatially
flat homogeneous and isotropic background cosmology given by the
metric
\begin{equation}
\mathrm{d}s^2 \, = \, -\mathrm{d}t^2 + a(t)^2 \mathrm{d}{\bf x}^2 \, ,
\end{equation}
where $t$ is physical time, ${\bf x}$ are the comoving spatial coordinates,
and $a(t)$ is the cosmological scale factor. The Hubble expansion rate is
\begin{equation}
H(t) \, \equiv \, \frac{{\dot a}}{a} \, .
\end{equation}
There are two coupling constants which appear in string theory.
The first is the
string coupling constant $g_s$ which measures the strength
of the string interactions. The value of $g_s$ is set by the
expectation value of the string theory dilaton field.  The
second coupling constant $\alpha^{`}$ measures the
strength of quantum effects. It is given by the square
of the string length.

\section{Review of Fibre Inflation}

Inflation driven by a scalar field has become the paradigm of
early universe cosmology (see e.g. \cite{Baumann1} for a review).
Unless the potential of the inflaton field is finely tuned, the
energy scale at which inflation takes place corresponds that
of particle physics Grand Unification, which is close to the
expected string scale in many setups \cite{GSW}.  Hence, it
is reasonable to search for realizations of inflation in the context
of string theory.

String theory is anomaly-free only in ten space-time dimensions.
Hence, to make contact with our world, six of the spatial dimensions
need to be compactified to a small size. In this context, the
effective field theory in our four space-time dimensions will
contain many scalar fields, some of which could be candidates
for the inflaton (see e.g. \cite{Baumann2} for a review of
the connection between inflation and string theory).

The scalar field candidates to be the inflaton include the
separation between branes \cite{Stephon, brane}, K\"{a}hler
moduli (roughly speaking the radii of the compactified dimensions)
\cite{Racetrack, Kahler}, and axions \cite{Witten, Keshav}.

Scalar field-driven inflation usually requires the field to be slowly
rolling (technically this means that the acceleration term in the
Klein-Gordon equation for the field is negligible, and the kinetic
energy is also negligible). These conditions are in general hard
to realize. In particular, if we want the slow-roll trajectory of the
scalar field to be a local attractor in initial condition space, then
field values greater than the Planck scale are required (see
e.g. \cite{ICreview} for a recent review and \cite{original} for
some original works). As reviewed in \cite{Baumann2}, such large
field values are hard to obtain in a controlled way from string
theory. There are, however, a couple of interesting possibilities,
the first being axion monodromy inflation \cite{axionmon},
the second the {\it Large Volume Scenario} \cite{LVS} \footnote{Both
of these scenarios may, in fact, lie in the string theory swampland \cite{Vafa},
but this is not the topic of our work.}.

Fibre inflation \cite{Cicoli} is a particular realization of inflation in the
context of the Large Volume Scenario. It is based on a flux compactification
of Type IIB superstring theory, with the compact manifold being
a Calabi-Yau three fold $X$ \cite{flux1, flux2} with fibres.
The resulting low energy theory is four space-time dimensional
${\cal N} =1$ supergravity which in turn is given by a K\"{a}hler potential $K$
(which determines the kinetic part of the action) and a superpotential $W$
which determines the potential energy term \footnote{The K\"{a}hler potential
also enters in the connection between the superpotential and the potential.}.

The K\"{a}hler potential depends on the total volume ${\cal{V}}$ of the
threefold, on the axio-dilaton $S$, on a holomorphic (3,0) form $\Omega$,
on complex structure moduli $U_{\alpha};  \,\, \alpha = 1, ... ,   h_{2,1}(X)$ and on
K\"{a}hler moduli $T_i ; \,\,  i  = 1, ... ,  h_{1,1}(X)$. Here the numbers
$h_{2,1}$ and $h_{1,1}$ characterize the topology of the compact space $X$.

The axio-dilaton field $S$ and the complex structure moduli are fixed by turning on fluxes
(both NS-NS and RR fluxes). At tree level the K\"{a}hler moduli $T_i$ are not fixed
(and neither is the volume ${\cal{V}}$, but the volume is determined
from the $T_i$ \footnote{Note that in the {\it string gas cosmology} \cite{BV2} approach
to merging string theory with cosmology (see \cite{SGCrev} for a review)
the K\"{a}hler moduli are automatically fixed \cite{Subodh} by the interplay of
winding and momentum modes. The winding modes are frozen out in the
effective field theory approach.}. There are
no corrections to the superpotential at any finite order in $g_s$ and $\alpha^{`}$.
However, there are non-perturbative corrections (e.g. via gaugino condensation),
and they give exponential terms of the form
\begin{equation}
\Delta W \, = \, \sum_i A_i e^{- a_iT_i} \, ,
\end{equation}
where $A_i$ are coefficients. The K\"{a}hler potential obtains perturbative
corrections of the form:
\begin{equation}
\Delta K \, = \, - 2 \ln \left( {\cal{V}} + \frac{\xi}{2 g_s^{3/2}} \right) \, ,
\end{equation}
where $\xi$ is a constant proportional to the Euler number of $X$.

With these corrections it is possible to generate a minimum of the potential
corresponding to large volume. But this depends on the geometry of $X$. The
Calabi-Yau manifold $X$ must have a ``blow-up'' mode (a direction in which
the radius is large in string units) to get the minimum of the potential
to be at a large value of ${\cal{V}}$. In this case, the
induced potential stabilizes the K\"{a}hler moduli in the directions which
are not large, but not the value of the blowup mode.
Note that extra structures are needed to uplift the AdS (anti-de-Sitter) minimum
to a dS (de Sitter) (and this might not be consistent from the point of view of string theory).

The specific example considered in \cite{Cicoli} has two blowup directions
which yield two light moduli fields $\tau_1$ and $\tau_3$. The overall
volume ${\cal{V}}$ is determined by the values of the blowup fields.
Without string loop corrections, the potential is flat in a direction which
in field space is roughly $\tau_1$. If we fix the total volume and $\tau_3$, then in fact
the flat direction is exactly the $\tau_1$ direction.
In $\tau_3$ direction there is a minimum of the potential, yielding a valley
in the potential viewed as a function of $\tau_1$ and $\tau_3$ which is the
inflaton direction. For values of $\tau_3$ larger than the value at
the minimum of the potential, the potential increases slowly. How
slowly will be the key question we ask in Section 4. String loop
effects will slightly lift the potential in $\tau_1$ direction, thus allowing
for inflation.

As done in \cite{Cicoli}, we here consider two versions of the fibre inflation
model. In the first we fix $\tau_3$ and ${\cal{V}}$ at the minima of their
potentials. We then obtain a single field inflation model given by the
Lagrangian
\be \label{Lag1}
{\cal L} \, = \,  - \frac{3}{8} \left( \frac{\partial_{\mu} \tau_1 \partial^{\mu} \tau_1}{\tau_1^2} \right)
- V(\tau_1)
\ee
with potential
\be \label{pot1}
V(\tau_1) \, = \, V_0 +
\left( \frac{A}{\tau_1^2} - \frac{B}{{\cal{V}} \sqrt{\tau_1}} + \frac{C \tau_1}{{\cal{V}}^2} \right)
\frac{W_0^2}{{\cal{V}}^2}
\ee
where $V_0, A, B, C$ and $W_0$ are constants given by the three-fold $X$,
and the volume ${\cal{V}}$ is taken to be constant.

In the second version we keep $\tau_3$
fixed at the minimum but allow ${\cal{V}}$ to fluctuate away from its minimum.
The kinetic piece of the Lagrangian is then both non-canonical and non-diagonal
\be \label{Lag2}
{\cal L}_{\text{kin}} \, = \,  -\frac{3}{8}  \left( \frac{\partial_{\mu} \tau_1 \partial^{\mu} \tau_1}{\tau_1^2} \right)
+ \frac{1}{2}  \left( \frac{\partial_{\mu} \tau_1 \partial^{\mu} {\cal{V}} }{\tau_1 {\cal{V}}} \right)
-  \frac{1}{2} \left( \frac{\partial_{\mu} {\cal{V}} \partial^{\mu} {\cal{V}}}{{\cal{V}}^2} \right)\, ,
\ee
and then potential is
\be \label{pot2}
V \, = \,  \left[ - \mu_4 ({\rm{ln}}(c {\cal{V}}))^{3/2} + \mu_3 \right] \frac{W_0^2}{{\cal{V}}^3}
+ \frac{\delta_{{\rm{up}}}}{{\cal{V}}^{4/3}}
+ \left( \frac{A}{\tau_1^2} - \frac{B}{{\cal{V}} \sqrt{\tau_1}} + \frac{C \tau_1}{{\cal{V}}^2} \right)
\frac{W_0^2}{{\cal{V}}^2} \,
\ee
where $c, \mu_4, \mu_3, W_0, \delta_{\rm{up}}, A, B$ and $C$ are constants
which are determined by the specific manifold $X$.

\section{Self-Resonance in the Single-Field Model}

Here we consider the model given by (\ref{Lag1}) and (\ref{pot1}). In order to
obtain the Lagrangian of a canonically normalized field we make the
field redefinition (recall that we are using units in which the Planck mass is set to $1$)
\be
\phi \, = \, \frac{\sqrt{3}}{2} {\rm{ln}} \tau_1 \, .
\ee
Then, the Lagrangian has the canonical form
\begin{equation}
\mathcal{L} \, = \, -\frac{1}{2}\partial_{\mu}\phi\partial^{\mu}\phi - V(\phi),
\end{equation}
where the potential of the inflaton is
\begin{equation}
V(\phi) \, = \, \frac{\mathcal{C}_2}{\langle\mathcal{V}\rangle^{10/3}} \left[(3-R)-4\left(1+\frac{R}{6}\right)e^{-\kappa\phi/2}+
\left(1+\frac{2R}{3}\right)e^{-2\kappa\phi}+R e^{\kappa\phi}\right].
\label{inflaton potential}
\end{equation}
where $R$ is a small parameter, $R\ll 1$, and $\kappa \equiv \frac{2}{\sqrt{3}M_p}$.

The homogeneous mode of the inflaton satisfies the Klein-Gordon equation
\begin{equation}
\ddot{\phi} + 3H\dot{\phi} + V_{\phi} \, = \, 0 \, ,
\label{field evolution}
\end{equation}
where the subscript indicates the argument with respect to which the derivative is taken.
For large field values $\kappa \phi > 1$ (but not $R {\rm{exp}}(\kappa \phi) > 1$) the
inflationary slow-roll conditions on the potential are satisfied. Hence, for such values
of $\phi$ inflation can take place. Once $\kappa \phi < 1$ inflation ends and $\phi$
begins to oscillate about the minimum of the potential. In the absence of cosmic
expansion (i.e. setting $H = 0$ in the above equations), the frequency
of the oscillations is $m_{\phi}$. Through the nonlinear term in the evolution
equation, self-resonance (i.e. resonance of fluctuations of the field $\phi$ itself)
is possible. However, to check whether this indeed takes place we need to take
into account the expansion of space.

\subsection{Evolution without cosmic expansion}

First we consider in more detail the evolution for $H = 0$.
Once $\kappa \phi < 1$, the slow-roll conditions break down, and the
inflaton rolls to the minimum of its potential and begins to oscillate about it.

\begin{figure*}[htb]
\begin{center}
\includegraphics[width=7cm,height=4.6cm]{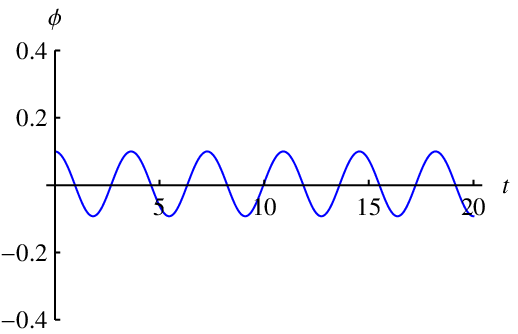}
\includegraphics[width=7cm,height=4.6cm]{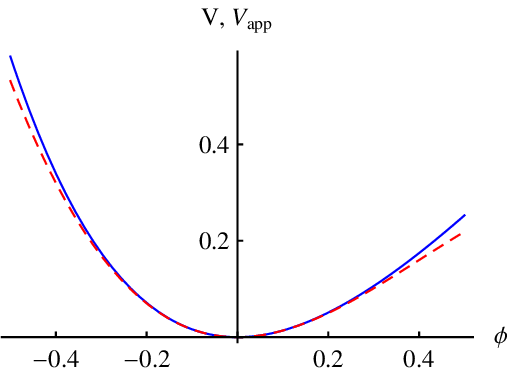}
\end{center}
\caption{The evolution (with $H=0$) of the homogeneous mode of the
inflaton (vertical axis) as a function of time (horizontal axis)
during reheating is shown in the left panel. The
right panel shows both the full and the approximated potential, the dashed red line
being $V_{\text{app}}$ and the blue line the full potential $V$.}
\label{Evolution_single_woE}
\end{figure*}

For small field amplitudes $\phi_0<1$ (in Planck units, $M_p=1$), the potential
can be approximated as
\begin{equation}
V_{\text{app}}(\phi) \, \simeq \, \frac{1}{2}m_{\phi}^2 \phi^2 + \frac{1}{6}g\phi^3,
\label{approximated potential}
\end{equation}
with
\ba
m_{\phi}^2 \, &=& \, \frac{\mathcal{C}_2(6+7R)\kappa^2}{2\langle\mathcal{V}\rangle^{10/3}} \label{mass} \\
g \, &=& \, -\frac{\mathcal{C}_2(17R+30)\kappa^3}{4\langle\mathcal{V}\rangle^{10/3}} \, .
\ea
Once the amplitude becomes small enough such that the effects of the nonlinear terms
in the equation of motion are negligible (this always happens since the inflaton
loses energy),  the oscillations become harmonic:.
\begin{equation}
\phi \, \simeq \, \phi_0 \, \text{cos} (\omega t) \, ,
\label{approximated solution}
\end{equation}
with $\omega=m_\phi$. We have denoted the amplitude of the homogeneous
mode by $\phi_0$.

The potential and a typical example of the evolution of the inflaton are shown in Fig. \ref{Evolution_single_woE}. The right panel shows the potentian and its approximated
version, the left panel shows the time evolution of $\phi$.

To study the self-resonance, let us consider the perturbation of the inflaton,
\begin{equation}
\delta\phi(x, t) \,  \equiv \, \phi(x, t) - \langle\phi\rangle(t) \, ,
\end{equation}
where the brackets indicate spatial averaging. To linear order,
the perturbation mode (in Fourier space) satisfies the equation of motion
\begin{equation}
\ddot{\delta\phi}_{\mathbf{k}}
+ \left(k^2 + V_{\phi\phi}\right)
\delta\phi_{\mathbf{k}} \, = \, 0 \, ,
\label{perturbation equation}
\end{equation}
where we omitted the Hubble damping and set $a = 1$.
For a large amplitude of oscillation, one should
calculate $V_{\phi\phi}$ with the complete potential (\ref{inflaton potential}) since the
higher order interactions are important in this case. As the amplitude decays to be small,
the approximation (\ref{approximated potential}) works well. We will consider this
case in what follows. With the solution (\ref{approximated solution}) we have
\begin{equation}
\frac{d^2V_{\text{app}}}{d\phi^2} \, = \, m_\phi^2 +  g \phi_0 \text{cos}(\omega t).
\end{equation}

Thus we have a Mathieu equation for the perturbation mode $\delta\phi_\mathbf{k}$,
\begin{equation}
\ddot{\delta\phi}_{\mathbf{k}}
+ \left(k^2+m_\phi^2 + g\phi_0 \text{cos}(\omega t)\right)
\delta\phi_{\mathbf{k}} \, = \, 0 \, .
\end{equation}
Using the rescaled time coordinate $\tau=\frac{1}{2}\omega t$, we get a standard Mathieu equation,
\begin{equation}
\delta\phi_{\mathbf{k}}''
+ \left[A_{\mathbf{k} }- 2q\text{cos}(2\tau)\right]
\delta\phi_{\mathbf{k}} \, = \, 0 \, ,
\label{mathieuWOE}
\end{equation}
where
\ba
A_{\mathbf{k}} \, &=& \, \frac{4(k^2+m_\phi^2)}{\omega^2} \nonumber \\
q \, &=& \, -\frac{2g\phi_0}{\omega^2} \, .
\ea

This equation has the following solution
\begin{equation}
 \delta\phi_{\mathbf{k}} \, = \, e^{{\tilde{\mu}}_k \tau}P_1(k,\tau)  + e^{-{\tilde{\mu}}_k \tau}P_2(k,\tau) \, ,
\end{equation}
where $P_{1,2}$ are periodic functions,
and ${\tilde{\mu}}_k$ is the famous Floquet exponent. If the Floquet exponent has nonvanishing
real part, then there is parametric resonance of the fluctuations \cite{TB}. In terms of physical
time we can write
\be
\mu_k t \, \equiv \, {\tilde{\mu}} \tau \, .
\ee
Note that $\mu_k$ has dimensions of energy, whereas ${\tilde{\mu}}$ is dimensionless.

To know how large the Floquet exponent is, let us analyze the parameters involved.
After the end of inflation, the field oscillates about the minimum of the potential, and the amplitude is small compared to the Planck scale. Hence, we can regard the background system as a harmonic oscillator. Combining the equations (\ref{approximated solution}) and (\ref{approximated potential}) we have $\omega\simeq m_\phi$. According the reference \cite{Cicoli}, we have
\ba
\kappa \, &=& \, \frac{2}{\sqrt{3}M_{p}} \nonumber \\
g \, &=& \, -\frac{5m_\phi^2}{\sqrt{3}M_p} \, .
\ea
We immediately get
\begin{equation}
A_\mathbf{k}\simeq \frac{4k^2}{m_\phi^2} + 4,\quad q\simeq \frac{10\phi_0}{\sqrt{3}M_p} \, .
\label{A and q without expansion}
\end{equation}

Since $\phi_0 < M_p$ (as discussed earlier), we can conclude that
\begin{equation}
A_{\mathbf{k}} > 4,\quad q < \frac{10}{\sqrt{3}} \, .
\end{equation}
\begin{figure*}[htb]
\begin{center}
\includegraphics[width=7cm,height=5.2cm]{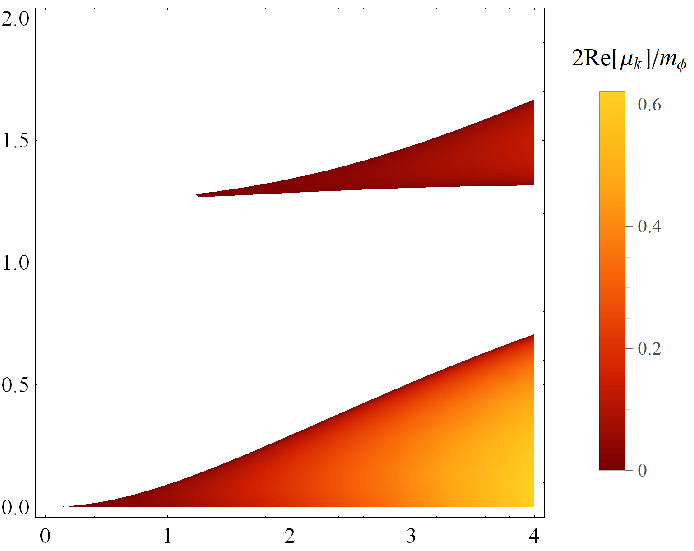}
\includegraphics[width=7.4cm,height=5.2cm]{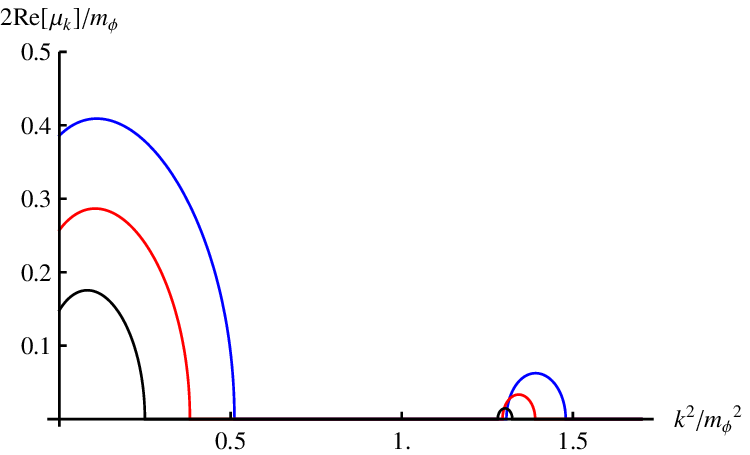}
\end{center}
\caption{The band structure for the Floquet exponent of equations (\ref{A and q without expansion}) and (\ref{mathieuWE}). On the left, the horizontal axis represents the parameter $q$, and the vertical axis is the parameter $k^2/m_\phi^2$, which represents the rescaled wavelength. The colored region is the parameter space where $\text{Re}[\mu_k]\neq 0$, with the color coding indicating the value of the Floquet exponent. The right panel shows the Floquet exponent for values of $k$ in the first parametric resonance band. The three curves are for $q=3$ (blue line), $q=2.4$ (red line), and $q=1.8$ (black line), which correspond to the inflaton oscillation with amplitude $\Phi=0.5$, $\Phi=0.4$, and $\Phi=0.3$ respectively.}
\label{bandstructure_Single_woE}
\end{figure*}
From the Fig. \ref{bandstructure_Single_woE}, we see that the rescaled Floquet exponent is smaller than $0.5$.
The first band is the most important one. This implies that the long wavelength modes dominate particle production.
Since the phase space of modes grows as $k^3$, the dominant modes are in fact the modes with a value of $k^2$ close to the upper end of the resonance band. Note that these modes are sub-Hubble (as can be seen by making use of the Friedmann equation), hence justifying the particle interpretation. The Floquet exponent for some typical values of $q$ is shown in Fig. \ref{bandstructure_Single_woE}.

To have an exact feeling about the efficiency of the self-resonance, we need to compare the Floquet exponent with the Hubble parameter $H$. If the former is larger than the latter, the exponential increase in the particle number is significant. We have
\begin{equation} \label{ratio}
\frac{\text{Re}[\mu_k]}{H} \, = \, \frac{m_\phi}{2H}\frac{2\text{Re}[\mu_k]}{m_\phi} \, ,
\end{equation}
which is larger than one. For the value of $m_{\phi}$ given by the parameters of the model (see \ref{mass}) we have $H \sim 10^{-1} m_{\phi}$ at the beginning of reheating, and with the maximal value of the rescaled Floquet exponent being about $0.5$, the ratio (\ref{ratio}) is about $5$. On the other hand, since the ratio is not significantly larger than one, it is important to extend our analysis and include the effects of the expanding background.

\subsection{Including the Effects of the Expanding Background}

\begin{figure*}[htb]
\begin{center}
\includegraphics[width=7cm,height=4.6cm]{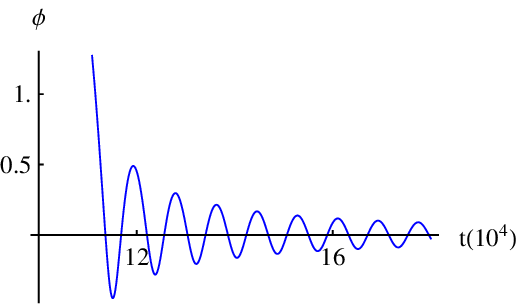}
\includegraphics[width=7cm,height=4.6cm]{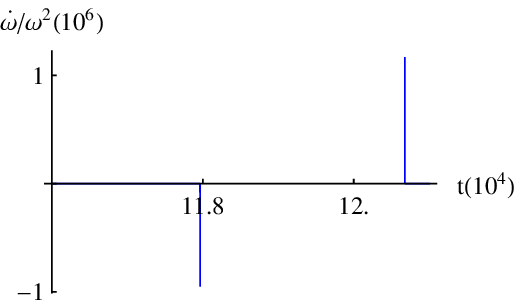}
\end{center}
\caption{The evolution of the inflaton and the violation of the adiabatic condition after inflation. The inflaton (vertical axis) oscillates after inflation as a function of time (horizontal axis), with the amplitude decaying and the periodicity being almost a constant.}\label{single_evolution wE}
\end{figure*}

Now let us include the cosmic expansion. In this case the evolution should be described by the equation (\ref{Evolution_single_woE}). For some given initial conditions, it has a typical evolution as in the Fig. \ref{single_evolution wE} (which is obtained by numericallly solving the evolution equation). As expected, the inflaton starts to oscillate about the minimum of the potential with an amplitude smaller than the Planck scale. Hence, it can be viewed as small perturbation about the minimum $\langle\phi\rangle$, and we can write
\begin{equation}
\delta\phi \, = \, \phi - \langle\phi\rangle \, .
\end{equation}

The perturbation mode (in Fourier space) satisfies the equation of motion
\begin{equation}
\ddot{\delta\phi}_{\mathbf{k}}+3H\delta\phi_{\mathbf{k}}
+\left(\frac{k^2}{a^2}+V_{\phi\phi}\right)
\delta\phi_{\mathbf{k}}=0.
\label{perturbation equation}
\end{equation}
It is convenient to eliminate the fraction term by making use of a field transformation
\begin{equation}
X_{\mathbf{k}} \, = \, a^{3/2}\delta\phi_{\mathbf{k}} \, .
\end{equation}
In terms of $X_{\mathbf{k}}$, the equation (\ref{perturbation equation}) becomes
\begin{equation}
\ddot{X}_{\mathbf{k}} + \left(\frac{k^2}{a^2}+V_{\phi\phi}-\frac{9}{4}H^2
-\frac{3}{2}\dot{H}\right) X_{\mathbf{k}} \, = \, 0 \, ,
\end{equation}
where
\be
\omega_{\mathbf{k}}^2(t) \, = \, \frac{k^2}{a^2}+V_{\phi\phi}-\frac{9}{4}H^2 -\frac{3}{2}\dot{H} \, .
\ee

This equation was studied in detail in \cite{KLS2}, where it was shown that particle production occurs when the adiabaticity condition on the time dependence of the frequency is violated. This happens when the background field $\phi$ takes on a local maximum (since then the Floquet exponent takes on a local minimum), as long as the amplitude of the background field remains sufficiently large.

We will give an approximate treatment (see e.g. \cite{STB} where it was shown that this analysis gives qualitatively the same results as the rigorous treatment of \cite{KLS2}) and set $a(t) = 1$ in the fluctuation equation. The equation for the perturbation mode then becomes
\begin{equation}
\ddot{X}_{\mathbf{k}} + \left(k^2 + V_{\phi\phi}\right) X_{\mathbf{k}} \, = \, 0 \, .
\label{mode equation}
\end{equation}
Now we compute $V_{\phi\phi}$, using the approximation (\ref{approximated potential}).
During the oscillations, the background field goes like $\phi(t)=\Phi(t)\text{cos}(\omega t)$ with $\Phi(t)\sim\frac{\lambda}{t}$, which is shown in Fig. \ref{single_evolution wE}. Note that the inflaton starts to oscillate when the amplitude $\Phi(t)$ is small, $\Phi(t) < 1$ (in Planck units), and the amplitude continues to decrease.The equation (\ref{mode equation}) then becomes
\begin{equation}
\ddot{X}_{\mathbf{k}} + \left(k^2 + m_{\phi}^2 + g\Phi(t)\text{cos}(\omega t)\right) X_{\mathbf{k}} \, = \, 0 \, .
\end{equation}
It is convenient to write it in terms of the new coordinate $\tau=\frac{1}{2}\omega t$, since  this yields the standard form of the Mathieu equation,
\begin{eqnarray}
{X}_{\mathbf{k}}'' + \left(A_{\mathbf{k}} -2 q \text{cos}(2\tau)\right) X_{\mathbf{k}} \, = \, 0 \, ,
\nonumber\\
A_{\mathbf{k}} \, = \, \frac{4k^2}{\omega^2}+\frac{4m_{\phi}^2}{\omega^2},\quad
q \, = \, -\frac{2g\Phi}{\omega^2}.
\label{mathieuWE}
\end{eqnarray}
where a prime represents the derivative with respect to $\tau$. As a good approximation, we can use $\omega\simeq m_\phi$. Then we find
\begin{equation}
A_{\mathbf{k}}\simeq\frac{4k^2}{m_\phi^2}+4,\quad
q\simeq \frac{10\Phi(t)}{\sqrt{3}M_p}.
\end{equation}
Typically, the inflaton oscillates for $\Phi(t) < 1$, hence $q < 6$. We see that in this case we have narrow resonance. Since the amplitude $\Phi(t)$ decays, the parameter $q$ becomes smaller and smaller, and this makes the resonance band narrower. From the Fig. \ref{bandstructure_Single_woE}, we conclude that the cosmic expansion reduces but does not cut off the efficiency of particle production.

\subsection{Termination of Preheating}

Due to the nonlinearities in the equation of motion for the background field, particle production will lead to correction terms. Once these correction terms become comparable to the mass term driving the oscillations, preheating will shut off. The leading nonlinearity comes from the cubic term in the approximate potential and yield the corrected equation of motion
\be
{\ddot{\phi}} + 3 H {\dot{\phi}} = - m_{\phi}^2 \phi - \frac{1}{2} g \phi^2 \, .
\ee
Each Fourier fluctuation mode $\delta \phi_k$ contributes to the correction term. The criteria for termination of preheating thus becomes
\be \label{crit}
m_{\phi}^2 \Phi_0 \, > \, g \langle(\delta \phi)^2\rangle \, ,
\ee
where, as before, $\Phi_0$ is the amplitude of the background field. The quantity $\langle(\delta \phi)^2\rangle$ is the contribution of the fluctuation modes to the background. It is given by
\be \label{fluct}
\langle(\delta \phi)^2\rangle \, = \, \int d^3k |\delta \phi_k|^2 \, ,
\ee
where $\delta \phi_k$ are the fluctuations computed earlier.

To compute this expectation value, consider the Fourier expansion of the fluctuation field
\be
\delta \phi \, = \, \int d^3k f_k \epsilon_k e^{i k x} \, ,
\ee
where $f_k$ is the amplitude of the fluctuation mode and $\epsilon_k$ is a random variable with normalization
\be \label{random}
\langle \epsilon_k \epsilon_{k'} \rangle \, = \, \delta^3(k - k') \, .
\ee
Since particle production is dominated by the first resonance band, we can estimate (\ref{fluct}) by restricting the integral to this first band and using
\be \label{amplitude}
f_k \, = \, \frac{1}{\sqrt{k}} e^{\mu_k t} \, ,
\ee
where we have assumed vacuum initial conditions for the amplitude of the Fourier modes. Inserting (\ref{random}) and (\ref{amplitude}) into (\ref{fluct}), we can then estimate the resulting integral by $k_{\star}^2$ (the upper cutoff value of the first resonance band) multiplied by the factor of increase, using the maximal value of the Floquet exponent in the first resonance band. This yields
\be
\langle(\delta \phi)^2\rangle \, \sim \, k_{\star}^2 e^{2 \mu_k t} \, ,
\ee
where $\mu_k$ can be taken to be the maximal value of the Floquet exponent.

The termination criterium (\ref{crit}) then becomes
\be \label{exponent}
2 \mu_k t \, <  \, {\rm{ln}}\bigl( \frac{m_{\phi}^2 \Phi_0}{g k_{\star}^2} \bigr) \, .
\ee
Since the oscillation period is given by $m_{\phi}^{-1}$ and since the maximal value of the Floquet exponent is about $0.5$ for $q = 0.7$, we find that the number of oscillation periods before the termination criterion is satisfied is given by the right hand side of the above equation. Making use of $k_{\star} \sim 10^{-1} m_{\phi}$ and that $g \simeq 3m_\phi^2/M_p$, we find that the number of oscillations is of the order $10$.

A better termination criterium is to check until which time adiabaticity violation takes place. As shown in Figure. \ref{single_evolution wE} (right), there is adiabaticity violation for a few oscillation times. This second analysis confirms the conclusion that the expansion of the universe and back-reaction do not prevent the preheating instability.

On the other hand, the preheating phase does not convert all of the energy of the initial inflaton condensate to particles. To estimate the fraction of the energy which is transferred during the initial stage of parametric resonance we can compute the energy density in particles, using
\be
\rho_{p} \, \sim \, 4 \pi \int_0^{k_{\star}} dk k^3 e^{2 \mu_k t} \, .
\ee
Inserting the result from (\ref{exponent}) and using $k_{\star} \sim 10^{-1} m_{\phi}$ we find
\be
\frac{\rho_p}{\rho_0} \, \sim \, 10^{-2} \frac{m_{\phi}^2}{g \Phi_0}
\ee
which is of the order of a percent.

In conclusion, in this section we have shown that parametric self-resonance is effective in the fibre inflation model with the model parameters suggested in \cite{Cicoli}. Note that in \cite{Antusch}, a much larger value of the Hubble constant was chosen. This choice plus small differences in the shape of the
potential lead to significant differences in the value of the Floquet
exponent. The authors of \cite{Antusch} obtain very similar results as
we do if they apply their analysis to the same form of the potential and parameter values as we use (private communication from F. Cefala). It is also important to stress, as done in \cite{Antusch}, that the initial presence of the preheating instability does not imply that this instability will drain most of the inflaton energy.

\section{Two-field Model}

\begin{figure*}[htb]
\includegraphics[width=7.345cm,height=4.5cm]{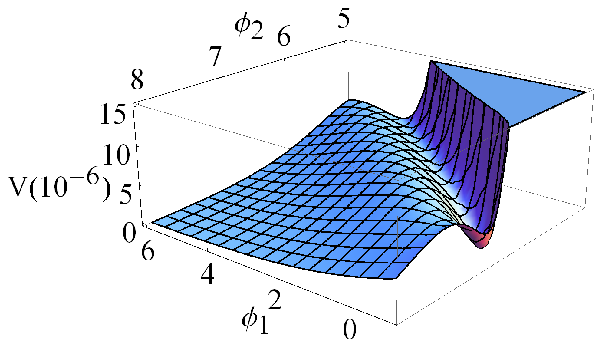}
\includegraphics[width=6.655cm,height=4.5cm]{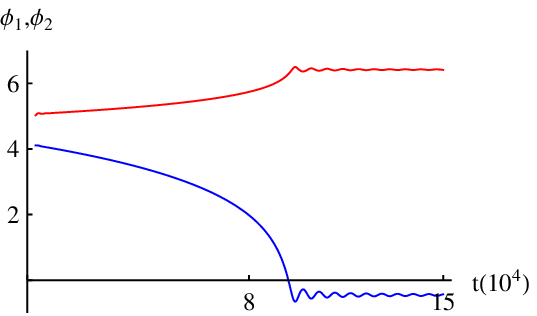}
\includegraphics[width=7cm,height=4.5cm]{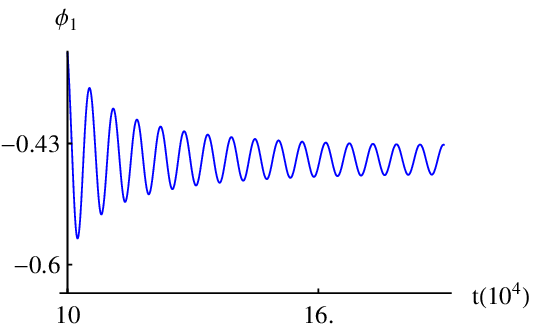}
\includegraphics[width=7cm,height=4.5cm]{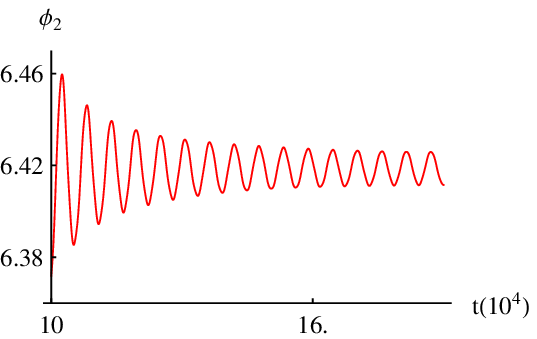}
\caption{The inflation potential $V(\phi_1,\phi_2)$ and the evolution of the scalar fields, with the blue line being $\phi_1(t)$ and the red line $\phi_2(t)$. We see that inflation mostly happens along the $\phi_1$. Note that we used the parameters of SV2 in \cite{Cicoli}.}
\label{evolution_twofields}
\end{figure*}

Now that we have shown that the initial stages of preheating occur via parametric self-resonance of inflaton fluctuations, we must study whether the parametric resonance might lead to entropy fluctuations of other light fields. In the fibre inflation model of \cite{Cicoli}, the Lagrangian including the second light degree of freedom is given by (see Section II)
\begin{equation}
\mathcal{L} \, = \, -\frac{3}{8\tau_1^2}\partial_\mu\tau_1 \partial^\mu\tau_1
+\frac{1}{2\tau_1 \mathcal{V}}\partial_\mu\tau_1 \partial^\mu\mathcal{V}
-\frac{1}{2\mathcal{V}^2}\partial_\mu\mathcal{V}\partial^\mu\mathcal{V} - V(\tau_1, \mathcal{V}),
\end{equation}
where the potential is
\begin{equation}
V(\tau_1, \mathcal{V}) \, = \, \left[-\mu_4(\ln(c\mathcal{V}))^{3/2}+\mu_3 \right]\frac{W_0^2}{\mathcal{V}^3}
+\frac{\delta_{up}}{\mathcal{V}^{4/3}}
+\left(\frac{A}{\tau_1^2}-\frac{B}{\mathcal{V} \surd \tau_1 }+\frac{C \tau_1}{\mathcal{V}^2}   \right)\frac{W_0^2}{\mathcal{V}^2}.
\end{equation}
We transform to canonically normalized fields $\phi_1$ and $\phi_2$ which are defined by
\begin{equation}
\tau_1 \, = \,  e^{a\phi_1 + b\phi_2},\quad
\mathcal{V} \, = \, e^{b\phi_1 + c\phi_2}.
\end{equation}
The dimensionless parameters above are $b=\sqrt{2/(7+4\sqrt{2})}$, $a=\sqrt{2-b^2}$, and $c=\sqrt{\frac{3}{2}-b^2}$. In terms of $\phi_1$ and $\phi_2$ one has the standard two-field theory
\begin{equation}
\mathcal{L} \, = \, -\frac{1}{2}\left(\partial_\mu\phi_1\partial^\mu\phi_1 +
\partial_\mu\phi_2\partial^\mu\phi_2\right) - V(\phi_1,\phi_2).
\label{twofield theory}
\end{equation}
Here the potential $V(\phi_1,\phi_2)$ is
\begin{eqnarray}
V(\phi_1,\phi_2) \, &=& \, -\mu_4W_0^2(\ln(c)+b\phi_1+c\phi_2)^{3/2}+\mu_3 W_0^2
e^{3b\phi_1+3c\phi_2}
+\delta_{up}e^{-4(b\phi_1+c\phi_2)/3}
\\ \nonumber
&+& \, A W_0^2 e^{-2(a+b)\phi_1-2(b+c)\phi_2}-B W_0^2 e^{-(\frac{1}{2}a+3b)\phi_1-\frac{1}{2}(b+3c)\phi_2}
+C W_0^2 e^{(a-4b)\phi_1+(b-4c)\phi_2} \, .
\label{two field potential}
\end{eqnarray}
The evolution of the background fields is described by the equations
\begin{equation}
\ddot{\phi}_i + 3H\dot{\phi}_i + V_{\phi_i} \, = \, 0 \, .
\end{equation}
The evolution of the two fields is shown in the Fig. \ref{evolution_twofields}.

In the following, we will show that in spite of the presence of two relevant field directions, entropy perturbations produced on super-Hubble scales during the preheating phase are too small to have an important effect of the spectrum of curvature perturbations. It turns ouit that the basic reason for this conclusion is that the second field is not sufficiently light compared to the Hubble expansion rate.

We will work in longitudinal gauge (see e.g. \cite{MFB} for a review of the theory of cosmological perturbations and \cite{RHBfluctRev} for an introductory overview) in which the metric, including scalar metric fluctuations, takes the form
\begin{equation}
ds^2 \, = \, -(1+2\Psi)dt^2 + a(t)^2(1-2\Phi)\delta_{ij}dx^i dx^j \, ,
\end{equation}
where $\Psi(x, t)$ and $\Phi(x, t)$ are the fluctuating degrees of freedom of the metric. If matter has no anisotropic stress (which is the case for scalar field matter), then one of the Einstein equations leads to the conclusion that the two potentials are equal. In this case, the curvature perturbation is
\be
\mathcal{R} \, = \, \Phi - \frac{H(H\Phi+\dot{\Phi})}{4\pi G(\rho+P)} \, .
\ee

In the absence of entropy fluctuations, the curvature perturbation $\mathcal{R}$ is conserved on super-Hubble scales. More generallly, the equation of motion for $\mathcal{R}$ is
\begin{equation}
\dot{\mathcal{R}} \, = \, -\frac{H}{\rho+P}\delta P_{nad}-\frac{H}{\rho+P}\frac{\dot{\rho}}{\dot{P}}
\frac{\nabla^2\Phi}{4\pi G a^2} \, ,
\end{equation}
where the non-adiabatic pressure perturbation $\delta P_{nad}$ is defined by
\begin{equation}
\delta P_{nad} \, = \, \delta P - \frac{\dot{P}}{\dot{\rho}}\delta\rho \, .
\end{equation}
Recalling the definition of the total entropy perturbation
\begin{equation}
\mathcal{S} \, \equiv \, H\left(\frac{\delta P}{\dot{P}}-\frac{\delta \rho}{\dot{\rho}}\right) \, ,
\label{total entropy}
\end{equation}
one immediately gets
\begin{equation}
\dot{\mathcal{R}} \, = \, -\frac{\dot{P}}{\rho+P}\mathcal{S}-\frac{H}{\rho+P}\frac{\dot{\rho}}{\dot{P}}
\frac{\nabla^2\Phi}{4\pi G a^2} \, .
\end{equation}
\begin{figure*}
\begin{center}
\includegraphics[width=7cm,height=5cm]{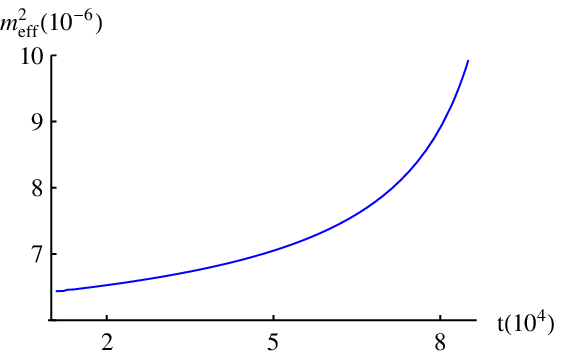}
\includegraphics[width=7cm,height=5cm]{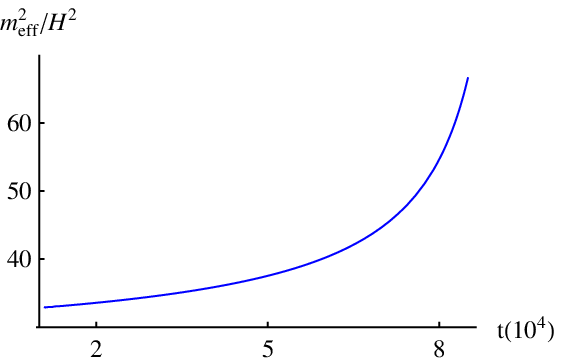}
\includegraphics[width=7cm,height=5cm]{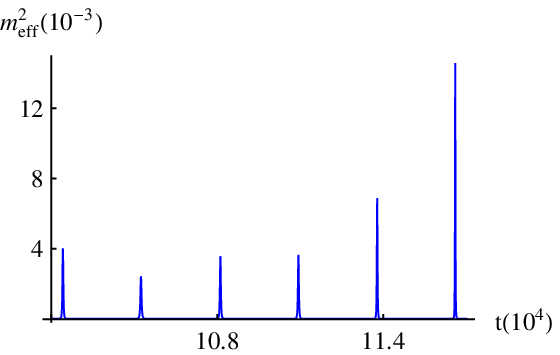}
\includegraphics[width=7cm,height=5cm]{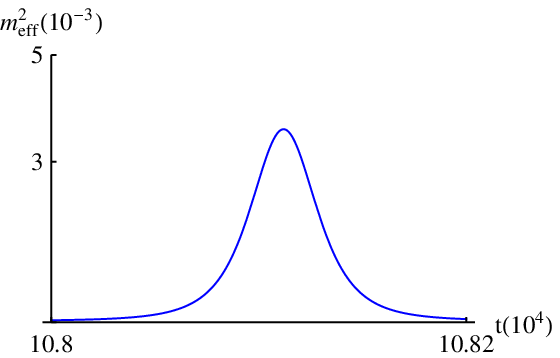}
\end{center}
\caption{The evolution of the effective mass of the entropy mode and its comparison to $H^2$.}
\label{effective mass}
\end{figure*}

On super-horizon scales, $k \ll aH$, the second term on the right-hand side can be ignored. Hence, on these large scales the evolution of the curvature perturbation is dominated by the entropy perturbation,
\begin{equation}
\mathcal{\dot{R}} \, \approx \, -\frac{\dot{P}}{\rho+P}\mathcal{S}\approx
-\frac{2\dot{\theta}\dot{P}}{3\dot{\sigma}(\rho+P)}\delta s.
\end{equation}
where $\delta s= \, \delta\phi_2\cos\theta-\delta\phi_1\sin\theta$, and the variables $\theta$ and $\sigma$ are defined in the Appendix.

The application of the above general formulas to two scalar field models has been studied in many references. We follow the treatment of \cite{Wands} which is summarized in the Appendix. The two fields are expanded into adiabatic field $\sigma$ and entropy field $s$. It is found that the entropy field fluctuation $\delta s$ obeys the equation
\be
\ddot{\delta s} + 3H\dot{\delta s} + \left(\frac{k^2}{a^2}+V_{ss}+3\dot{\theta}^2\right)\delta s \, = \,
\frac{\dot{\theta}}{\dot{\sigma}}\frac{k^2}{2\pi G a^2}\Phi \, ,
\label{entropy equation}
\ee
where the subscripts on $V$ indicate that the second derivative with respect to the entropy field is taken. The variable $\theta$ describes the angle between the original field basis and the adiabatic/entropy field basis. We see that the effective mass of the entropy mode is given by
\begin{equation}
m_{\text{eff}}^2 \, = \, V_{ss}+3\dot{\theta}^2 \, .
\label{definition of effective mass}
\end{equation}

For the two-field theory (\ref{twofield theory}), we have used the numerically determined background trajectory to determine the angle $\theta$ and the second derivative of the potential with respect to the entropy field. In this way, we have numerically evaluated the above effective mass. The results for the evolution of the effective mass square $m_{\text{eff}}^2$ are shown in Fig. \ref{effective mass}.
We see that the effective mass is much larger than $\frac{3H^2}{2}$, so the entropy mode is heavy. Hence, we expect the entropy mode to undergo damped oscillations. The top two panels show the evolution towards the end of the period of slow-roll inflation, the bottom two during the period of oscillations of the inflaton field about the minimum of its potential. The top left panel shows the evolution of the square of the mass, the top right the ratio of this square mass to the square of the Hubble expansion rate. The two bottom panels show a different range on the horizontal axis. With the scaling on the left we can track the overall amplitude, As is apparent, the oscillations of the inflaton lead to large amplitude peaks in the effective mass. However, the amplitudes vary dramatically from peak to peak, although the distance in time between the peaks appears to be nearly periodic. Basically, this is because of the structure of the potential (\ref{two field potential}). When expanded around the minimum for small amplitudes of oscillations, the potential has a crossed mass term for $\phi_1$ and $\phi_2$ and this breaks the harmonic nature of the oscillations. Since the time dependence of the effective square mass is far from periodic, there is no Floquet theory (parametric resonance) instability.

\begin{figure*}
\begin{center}
\includegraphics[width=7cm,height=5cm]{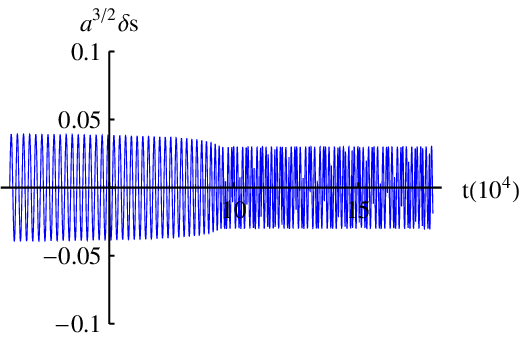}
\includegraphics[width=7cm,height=5cm]{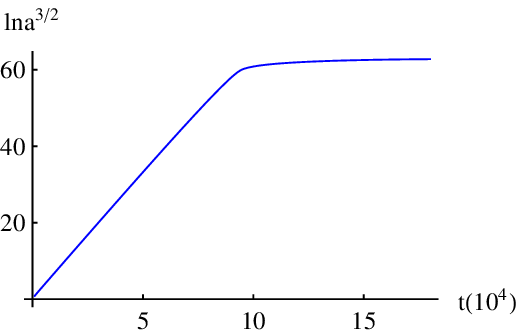}
\end{center}
\caption{The evolution of $a^{3/2}\delta s$ and the scale factor $a(t)$ during inflation and reheating. Note that $t=0$ is just the starting point of our numerical evolution but not of inflation, so we do not show the full 60 e-foldings.}
\label{entropy solution}
\end{figure*}

Indeed, given the numerically computed mass term we can numerically solve the entropy mode equation. The result is shown in Figure (\ref{entropy solution}).
To extract the effects of the expansion of space, we first make a transformation of variables, $\delta\tilde{s}=a^{3/2}\delta s$. Then the equation (\ref{entropy equation}) for the entropy mode becomes
\begin{equation}
\ddot{\delta\tilde{s}}+\left(V_{ss}+3\dot{\theta}^2-\frac{9}{4}H^2
-\frac{3}{2}\dot{H}\right)\delta\tilde{s}=0
\end{equation}
on large scales where gradient terms are negligible. Since the effective square mass is large compared to $H^2$, the $H^2$ and ${\dot{H}}$ terms in the equation are negligible. We have neglected them in the
numerical solution for $\delta\tilde{s}$ shown in Fig. \ref{entropy solution}.
We see that $\delta\tilde{s}$ oscillates with almost constant amplitude during inflation. In the reheating process, the amplitude has both amplification and decay. This can be understood by looking at the evolution of the effective mass in Fig. \ref{effective mass}. The effective mass of the entropy mode does not change much during inflation and this results in a nearly constant amplitude of $\delta\tilde{s}$. As mentioned above, during reheating, the effective mass has periodic narrow peaks. Except during the narrow peak intervals, the effective mass is almost constant, and no growth of $\delta\tilde{s}$. Since the peak structures are not periodic, they do not allow for the self-resonance for the entropy mode. Since $\delta s=a^{-3/2}\delta\tilde{s}$, considering the expansion of the universe, we conclude that the entropy mode is suppressed.

Note that in this analysis we have assumed that the background scalar field dominates the cosmology  during reheating. This is a good approximation at the beginning of the reheating phase, but will no longer be good once backreaction of the produced particles becomes important.

Since the entropy fluctuations do not grow, the induced effect they have on curvature fluctuations is expected to be negligible. We can check this by solving the equation (\ref{adiabatic equation}) for the adiabatic mode  (which gives the curvature perturbation via $\mathcal{R}=\frac{H}{\dot\sigma}Q$) given the entropy source function which we have determined. Like in the case of the entropy mode, we first extract the effects of the expansion of space via the transformation $Q=a^{-3/2}\widetilde{Q}$. This rescaled field obeys the equation
\begin{eqnarray}
\ddot{\widetilde{Q}}+\left(\frac{k^2}{a^2}+V_{\sigma\sigma}-\dot{\theta}^2
-8\pi G \left(3\dot{\sigma}^2+
\frac{2\dot{\sigma}\ddot{\sigma}}{H}-\frac{\dot{\sigma}^2\dot{H}}{H^2}\right)
-\frac{9}{4}H^2-\frac{3}{2}\dot{H}\right)\widetilde{Q}
\\ \nonumber
= \, 2\left(\ddot{\theta}-\frac{3}{2}\dot{\theta}H-\frac{V_\sigma}{\dot{\sigma}}
-\frac{\dot{H}}{H}\right)\delta\tilde{s}+2\dot{\theta}\delta\dot{\tilde{s}}
\end{eqnarray}

\begin{figure*}
\begin{center}
\includegraphics[width=7cm,height=5cm]{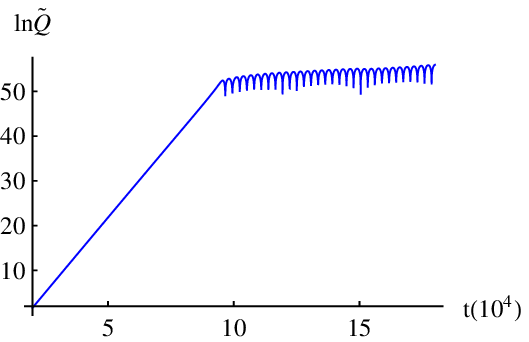}
\includegraphics[width=7cm,height=5cm]{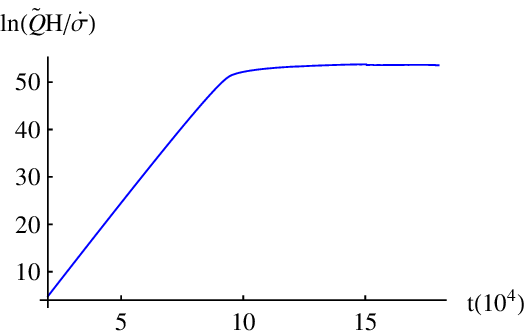}
\end{center}
\caption{The evolution of the mode $\widetilde{Q}$ and $\ln\left(\frac{H}{\dot{\sigma}}\widetilde{Q}\right)$ during inflation and reheating.}
\label{adiabatic solution}
\end{figure*}

The Evolution of the adiabatic mode is shown in Fig. \ref{adiabatic solution}.
The left plot shows the evolution of $\widetilde{Q}$. Combining the evolution of the scale factor in Fig. \ref{entropy solution}, we see that the adiabatic mode $Q=a^{-3/2}\widetilde{Q}$ does not grow. Actually, since
\begin{equation}
\ln \mathcal{R}=\ln\left(\frac{H}{\dot{\sigma}}\widetilde{Q}\right)
-\frac{3}{2}\ln a + \text{const} ,
\end{equation}
from the plot of $\ln\left(\frac{H}{\dot{\sigma}}\widetilde{Q}\right)$ in Fig. \ref{adiabatic solution} (right) we conclude that the curvature perturbation $\mathcal{R}$ is a constant during inflation and reheating.

\section{Conclusions}

We have studied the reheating phase in fibre inflation, taking the parameters used in \cite{Cicoli}. In this model, there are two relevant ``light'' degrees of freedom. They come from the moduli fields of the model. In a first approximation, we (following \cite{Cicoli}) have focused on the modulus field which taken by itself can lead to inflation. We have shown that the reheating phase in this model begins with a phase of parametric self-resonance during which a not negligible fraction of the inflaton energy is transferred to low momenta quanta of the inflaton.

We then considered the effects of the second lightest field. In order for the model to be safe, it is important to check that there is no strong parametric resonance of the fluctuations in this second field. Otherwise, the induced entropy fluctuations might lead to a contribution to curvature fluctuations which would destroy the successful predictions of the model. We have shown that, although the second field is light compared to the string scale, it is sufficiently heavy compared to the Hubble scale at the end of inflation such that super-Hubble entropy fluctuations cannot grow.

We have chosen the parameters of fibre inflation suggested in \cite{Cicoli}. It would be interesting to scan a wider parameter space of fibre inflation models and search for regions where there is a parametric instability of the entropy mode.

\section{Acknowledgement}

We thank Ryo Namba, Ziwei Wang, and Daisuke Yoshida for helpful discussions. In particular, we thank F. Cefala for useful communications concerning
the relation of our work to that of \cite{Antusch}.
B.-M. G. is supported by the Fundamental Research Funds for the Central Universities (Grants No. lzujbky-2017-it69) and by a scholarship granted by the China Scholarship Council (CSC, No. 201706180072).
The research at McGill is supported in
part by funds from NSERC and from the Canada Research Chairs program.

\section{Appendix}

Here we review the analysis of entropy perturbation $\mathcal{S}$ and induced curvature fluctuations in the case of two scalar field matter models, following the analysis of \cite{Wands}.

For scalar field matter (fields with canonical kinetic terms), the total energy density $\rho$ and pressure $P$ are given by
\be
\rho \, = \,\sum_I \frac{1}{2}\dot{\phi}_I^2 + V
\ee
and
\be
P \, = \, \sum_I \frac{1}{2}\dot{\phi}_I^2 - V \, ,
\ee
where $V$ is the potential energy function and the sum runs of the different fields. The time
derivatives and fluctuations of these functions are given by
\ba
\dot{\rho} \, &=& \, \sum_I \left(\dot{\phi}_I\ddot{\phi}_I +V_{\phi_I}\dot{\phi}_I\right),
\\
\dot{P} \, &=& \, \sum_I \left(\dot{\phi}_I\ddot{\phi}_I -V_{\phi_I}\dot{\phi}_I\right),
\\
\delta{\rho} \, &=& \, \sum_I\left(\dot{\phi}_I\delta\dot{\phi}_I
-\dot{\phi}_I^2\Phi\right)+\delta V,\label{density perturbation}
\\
\delta P \, &=&\, \sum_I\left(\dot{\phi}_I\delta\dot{\phi}_I
-\dot{\phi}_I^2\Phi\right)-\delta V.
\ea
Using these quantities, the total entropy perturbation (\ref{total entropy}) turns to be
\begin{equation}
\mathcal{S} \, = \, \frac{2\left(\dot{V}+3H\sum_I\dot{\phi}_I^2\right)\delta V
+2\dot{V}\sum_I\left(\dot{\phi}_I\delta\dot{\phi}_I-\dot{\phi}_I^2\Phi\right)}
{3\sum_I\dot{\phi}_I^2 \sum_J \left(2\dot{V}+3H\sum_J\dot{\phi}_J^2\right)} \, .
\end{equation}
If there is only one scalar field, then
\begin{eqnarray}
\mathcal{S} \, &=& \, \frac{2V_\phi}
{3\dot{\phi}  \left(2\dot{V}+3H\dot{\phi}^2\right)}
\left(\dot{\phi}\delta\dot{\phi}
-\dot{\phi}^2\Phi-\ddot{\phi}\delta\phi\right)
\nonumber\\
&=&\, \frac{2V_\phi}
{3\dot{\phi}  \left(2\dot{V}+3H\dot{\phi}^2\right)}
\left(\delta\rho+3H\dot{\phi}\delta\phi\right) \, .
\end{eqnarray}
In this case, the total entropy perturbation can be reduced by using the perturbed Poisson equation (the $00$ component of the Einstein equation) and the momentum constraint (the $0i$ component of the perturbed Einstein equation),
\begin{eqnarray}
\delta\rho + 3H\dot{\phi}\delta\phi\, = \, \frac{\nabla^2\Phi}{4\pi G a^2}.
\end{eqnarray}
The total entropy perturbation hence scales as $\mathcal{S}\propto \frac{k^2}{a^2}\Phi$ and is hence suppressed on super-Hubble scales. Hence, it is clear that the entropy perturbations vanish on large scales for single scalar field inflation models.

For two field models there is a convenient field space decomposition. The original scalar fields $\phi_1$ and $\phi_2$ can be described by the adiabatic field along the inflationary trajectory, and the entropy field orthogonal to the trajectory. The adiabatic field and its perturbation are
\begin{equation}
\dot{\sigma} \, = \, \dot{\phi}_1 \cos\theta+\dot{\phi}_2 \sin\theta,\quad
\delta\sigma \, = \, \delta\phi_1 \cos\theta+\delta\phi_2 \sin\theta \, ,
\end{equation}
where the angle $\theta$ is defined by
\begin{equation}
\cos\theta \, = \, \frac{\dot{\phi}_1}{\sqrt{\dot{\phi}_1^2+\dot{\phi}_2^2}},\quad
\sin\theta \, = \, \frac{\dot{\phi}_2}{\sqrt{\dot{\phi}_1^2+\dot{\phi}_2^2}}.
\end{equation}
The entropy field simply vanishes along the background trajectory.

The entropy perturbation is
\begin{equation}
\delta s \, = \, \delta\phi_2 \cos\theta-\delta\phi_1 \sin\theta \, .
\end{equation}
Using the density perturbation (\ref{density perturbation}) and the momentum constraint ($0i$ component of the Einstein equation) one can show that
\begin{eqnarray}
\mathcal{R} \, &=&\, \Phi+\frac{H}{\dot{\sigma}}\delta\sigma,
\nonumber\\
\mathcal{S} \, &=& \, -\frac{V_\sigma}{6\pi G \dot{\sigma}^2
\left(3H\dot{\sigma}+2V_\sigma\right)}\left(\frac{k^2}{a^2}\Phi\right)
+\frac{2\dot{\theta}}{3\dot{\sigma}}\delta s \, .
\end{eqnarray}
To know how large the entropy perturbation is, and its contribution to the curvature perturbation, it is necessary to know the evolution of the entropy perturbation mode $\delta s$ and the adiabatic mode $\delta\sigma$. The evolution equations of the entropy mode and the adiabatic mode are
\begin{eqnarray}
\ddot{\delta s}&+&3H\dot{\delta s} + \left(\frac{k^2}{a^2}+V_{ss}+3\dot{\theta}^2\right)\delta s \, = \,
\frac{\dot{\theta}}{\dot{\sigma}}\frac{k^2}{2\pi G a^2}\Phi,
\label{entropy equation}
\\
\ddot{Q}&+&3H \dot{Q}+\left(\frac{k^2}{a^2}+V_{\sigma\sigma}-\dot{\theta}^2
-\frac{8\pi G}{a^3}\partial_t \left(\frac{a^3\dot{\sigma}^2}{H}\right)\right)Q \,
= \, 2\partial_t
(\dot{\theta}\delta s)
-2\left(\frac{V_{\sigma}}{\dot{\sigma}}+\frac{\dot{H}}{H}\right)
\dot{\theta}\delta s,
\label{adiabatic equation}
\end{eqnarray}
where
\be
Q \, \equiv \, \delta\sigma+\frac{\dot{\sigma}}{H}\Phi \, .
\ee

Now we see that the effective mass of the  entropy mode is
\begin{equation}
m_{\text{eff}}^2 \, = \, V_{ss}+3\dot{\theta}^2 \, .
\label{definition of effective mass}
\end{equation}
The effective mass can be expressed in terms of the slow-roll parameters,
\begin{equation}
\varepsilon_{I J}=\frac{M_{pl}^2}{2}\frac{V_{\phi_I} V_{\phi_J}}{V^2},\quad
\eta_{IJ}=M_{pl}^2\frac{V_{\phi_I \phi_J}}{V},
\end{equation}
where
\be
V_{\phi_I} \, = \, \frac{\partial V}{\partial \phi_I} \,\,\,  {\rm{and}} \,\,\,
V_{\phi_I \phi_J} \, = \, \frac{\partial^2 V}{\partial \phi_I \partial \phi_J} \, .
\ee
These are the slow-roll parameters in the original field space. In the adiabatic and entropy field space, one has
\begin{eqnarray}
V_\sigma&=&V_{\phi_1}\cos\theta +V_{\phi_2}\sin\theta ,
\\
V_s&=&V_{\phi_2}\cos\theta -V_{\phi_1}\sin\theta,
\\
V_{\sigma\sigma}&=&V_{\phi_1\phi_1} \cos^2 \theta+ 2V_{\phi_1\phi_2}\sin\theta\cos\theta+V_{\phi_2\phi_2}\sin^2\theta,
\\
V_{s s}&=&V_{\phi_2\phi_2} \cos^2 \theta- 2V_{\phi_1\phi_2}\sin\theta\cos\theta+V_{\phi_1\phi_1}\sin^2\theta,
\\
V_{\sigma s}&=&-V_{\phi_1\phi_1} \sin\theta\cos\theta+ V_{\phi_1\phi_2}(\cos^2\theta-\sin^2\theta)
+V_{\phi_2\phi_2}\sin\theta\cos\theta.  \, .
\end{eqnarray}
One can show that
\begin{eqnarray}
\left(
  \begin{array}{cc}
    V_{\phi_1\phi_1} & V_{\phi_1\phi_2} \\
    V_{\phi_2\phi_1} & V_{\phi_2\phi_2} \\
  \end{array}
\right)=
 \left(
  \begin{array}{cc}
            \cos\theta & -\sin\theta \\
            \sin\theta & \cos\theta \\
  \end{array}
 \right)
\left(
  \begin{array}{cc}
    V_{\sigma\sigma} & V_{\sigma s} \\
    V_{s \sigma} & V_{s s} \\
  \end{array}
\right)
\left(
  \begin{array}{cc}
            \cos\theta & \sin\theta \\
           -\sin\theta & \cos\theta \\
  \end{array}
 \right)
\end{eqnarray}
Similarly, one has
\begin{eqnarray}
\left(
  \begin{array}{cc}
    \eta_{11} & \eta_{12} \\
    \eta_{21} & \eta_{22} \\
  \end{array}
\right)=
 \left(
  \begin{array}{cc}
            \cos\theta & -\sin\theta \\
            \sin\theta & \cos\theta \\
  \end{array}
 \right)
\left(
  \begin{array}{cc}
    \eta_{\sigma\sigma} & \eta_{\sigma s} \\
    \eta_{s \sigma} & \eta_{s s} \\
  \end{array}
\right)
\left(
  \begin{array}{cc}
            \cos\theta & \sin\theta \\
           -\sin\theta & \cos\theta \\
  \end{array}
 \right)
\end{eqnarray}
Hence the result is $V_{ss}=3H^2 \eta_{s s}$. For $\dot{\theta}$, we have
\begin{equation}
\dot{\theta} \, = \, -\frac{V_s}{\dot{\sigma}} \, = \, \frac{\dot{\phi}_1 \ddot{\phi}_2-
\dot{\phi}_2 \ddot{\phi}_1}{\dot{\sigma}^2}.
\end{equation}
Using the slow-roll equations it can be shown that
\begin{equation}
\dot{\theta} \, = \, -H \eta_{\sigma s} \, .
\end{equation}
Hence, we finally have
\begin{equation}
m_{\text{eff}}^2 \, = \, 3H^2(\eta_{s s}+\eta_{\sigma s}^2).
\label{slowrollmass}
\end{equation}
If the effective mass of the entropy mode is too heavy (larger than $\sqrt{\frac{3}{2}}H$), then the fluctuations coming from the entropy mode do not grow on large scales. Instead, they perform damped oscillations. Now let us compare this with the quantity $\frac{3}{2}H^2$,
\begin{equation}
m_{\text{eff}}^2-\frac{3}{2}H^2 \, = \, 3\left(\eta_{s s}+\eta_{\sigma s}^2-\frac{1}{2}\right)H^2 \, .
\label{comparison of mass}
\end{equation}
This implies that if $\eta_{s s}+\eta_{\sigma s}^2>\frac{1}{2}$ then the entropy perturbation are suppressed on large scales, Otherwise, the entropy perturbation can grow and give a significant contribution to the curvature perturbations. However, one should note that this analysis is only applicable during inflation since the equation (\ref{comparison of mass}) is obtained by using the slow-roll condition. For the evolution after slow-rolling, we should use the expression (\ref{definition of effective mass}) rather than (\ref{slowrollmass}).

\end{document}